
\input harvmac

\Title{\vbox{\baselineskip12pt\hbox{INRNE-1/93}}}
{\vbox{\centerline{Chern-Simons $p$-Branes and $p$-Dimensional}
   \vskip2pt\centerline{Classical  $W$-Algebras}}}

\centerline{Raiko P. Zaikov\footnote{$^\star$}
{Supported by Bulgarian Fondation on Fundamental Research under
contract Ph-11/91-94; \ \ \ \ e-mail zaikov@bgearn.bitnet}}
\bigskip\centerline{Institute for Nuclear Research and Nuclear Energy}
\centerline{Boul. Tzarigradsko Chaussee 72, 1784 Sofia}

\vskip .3in

\noindent{\bf Abstract.} It is shown that the generalized (with
nonpolynomial Lagrangian) Chern-Simons membranes and in general
$p$-branes moving in $D$-dimensional target space admit an infinite
set of secondary constraints. With respect to the Poisson bracket
these constraints satisfy closed algebra containing
$p$-dimensional classical $W$ algebra as a subalgebra. In the
case when the target space dimension $D$ is finite the theory is
topological.

\Date{3/93} 


\noindent The polynomial Chern-Simons $p$-brane action was
obtained \ref\rZa{R. P. Zaikov, Phys. Lett., {\bf B 266} (1991)
303} from the topological $(p+1)$-brane action
\ref\rF{B. Biran, E. G. Floratos and G. K. Savvidi, Phys. Lett.
{\bf B 198} (187) 329}\nref\rZ{R. P. Zaikov, Phys. Lett. {\bf B
211} (1988) 281}\nref\rGTz{M. P. Grabowski and C-H. Tze, Phys.
Lett. {\bf B 224} (1989) 259}
--\ref\rFKT{K. Fudjikawa, J. Kubo and H. Terao,
Phys. Lett., {\bf B 263} (1991) 371.} in the same way as in the
ordinary local theory.  In the paper \ref\rZa{R. P. Zaikov, Phys.
Lett., {\bf B 263} (1991) 209} generalization for arbitrary
space-time dimension was found by passing to nonpolynomial
Lagrangian in a similar form as the Nambu-Goto Lagrangian in the
ordinary string theory.  The Chern-Simons (C-S) string was
considered in
\ref\rSZa{M. N. Soilov and R. P. Zaikov, Lett. Math. Phys., {\bf
27} (1993) 155} where it was shown that two possibilities to have
first class constraints can be choisen: the first one is to have
a finite number of secondary constraints and second one is to
have an infinite number of secondary constraints. There is also
another possibility when second class constraints appear (see
\rZa \ and
\ref\rSZ{M. N. Stoilov and R. P. Zaikov, Mod. Phys. Lett., {\bf 8}
(1993) 313}).

In the case when the infinite set of secondary constrains apears
for the C-S string these constrains satisfy an infinite algebra
with respect to the Poisson brackets which contains as a
subalgebra the classical (without central term) $SL(2,R)$
Kac-Moody algebra \ref\rK{V. G. Kac, Func. Anal. Appl., {\bf 1}
(1967) 328; R. V.  Moody, Bull. Am. Math. Soc., {\bf 73} (1967)
328}, \ref\rBH{K.  Bardaci and M. B. Halpern, Phys. Rev., {\bf
D3} (1971) 2493} (complete list of references is given in
\ref\rHO{{\it "Ward Identities for Affine-Virasoro Corelators"},
Preprint LBL-32619; UCB-PTH-92/24; BONN-HE-92/21}) as well as the
classical Virasoro algebra and their higher spin extensions
containing classical $W_{1+\infty }$ algebra \ref\rP{C. N. Pope,
L. J. Romans and X. Shen, Pys. Lett. {\bf B 336} (1990) 173 and
{\bf 242B} (1990) 401.}.

In the present article the C-S membranes are considered.
Starting from the Nambu-Goto type C-S action with nonpolynomial
Lagrangian it is shown that when we deal with first class
constraints, always appears an infinite set of secondary
constraints. The latter is one of the differences from the C-S
string theory. These constraints give a linear realization of
two-dimensional higher spin extended algebra containig as
subalgebras two-dimensional $SL(2,R)$ Kac-Moody algebra
two-dimensional Virasoro algebra and two dimensional $W_{1+\infty
}$-algebra\foot{In ordinary two-dimensional conformal theories we
have two copies of Kac-Moody, Virasoro and $W$ algebras each of
which acts only on one light-cone (holomorphic or
antiholomorphic) coordinate, i.e. we have direct product of two
one-dimensional algebras.} .  We note that any of these algebras cannot be
represented
as a direct product of one-dimensional algebras as in
two-dimensional conformal theory
\ref\rFR{J. M. Figueroa-O'Farrill and E. Ramos, Phys. Let., {\bf
B 282} (1992) 357.}
\nref\rMR{F. Martinez-Moras and E. Ramos, {\it Higher Dimensional
Classical $W$-algebras}, Preprint-KUL-TF-92/19 and US-FT/6-92,
(hep-th/9206040)}--\ref\rMMR{F. M. Moras, J. Mas and E. Ramos,
{\it Diffeomorphisms from higher dimensional $W$-algebras},
Preprint, QMW-PH-93-1, US-FT/1-91, (hep-th/9303034)}\foot{In the
present article we use the terminology of the papers \rMR \ and
\rMMR .}. The phase space transformation laws are also
given. The generalization of the results for arbitrary C-S
$p$-branes is strigtforward. In that case we have higher spin
extension of the $p$-dimensional $SL(2,C)$ Kac-Moody algebra.

We start with the p-brane Nambu-Goto action
\eqn\eaa{S=\kappa \int _{\cal M}d\tau d^p\sigma \sqrt {-g},}
where
\eqn\eaaa{g=detg_{ab}}
\eqn\eab{g_{ab}=X^\mu _{,a}X^\nu _{,b}\eta _{\mu \nu },\qquad (\mu
,\nu =0,1,\dots ,D-1), \qquad (a,b=0,1,\dots ,p),}
$\eta _{\mu \nu }$ is the metric tensor in $D$ dimensional Minkowski
space-time, $X^\mu _{,a}=\partial _aX^\mu (a=0,1,\dots ,p)$,
$\kappa $ is $p$-brane tension and $\cal M$ is $p+1$-dimensional
area in the $D$-dimensional target space. The action \eaa \ is
invariant with respect to the $p+1$-variable diffeomorphisms and
as a consequence we have $p+1$ first class constraints:
\eqn\eac{\eqalign{\phi _{\perp }&={\cal P}^2+\kappa ^2det(g_{jk}),\cr
\phi _j&={\cal P}X_{,j}, \qquad \qquad \qquad (j,k=1,2,\dots ,p)\cr}}
where $g_{jk}$ are the spatial components of the induced metric
tensor \eab .

Now, let us consider the special case $D=p+1$. For example, in the
particle case ($p=0$) $D=1$, in the strings case ($p=1$) $D=2$,
in the membrane case ($p=2$) $D=3$, etc.. In that case
$X^\mu_{,a}$ becomes a square matrix and consequently the formula
\eaaa \ can be rewriten in the form:
\eqn\ead{detg=\Bigl(detX^\mu _{,a}\Bigr)^2.}
The equation \ead \ allows us to rewrite the action \eaa \ in the
following form
\eqn\eae{S=\kappa \int _{\cal M}d\tau d^p\sigma \vert
X^\mu_{,a}\vert .}
We note, that the Lagrangian in \eae \ is just the Jacobian of
the diffeomorphism which carries the world-sheet space into
$p+1$-dimensional target space and vice-versa. Consequently, in
that case all the components of $X^\mu $ are purely longitudinal,
i.e.  we have no transversal components. The topological
character of the action \eae \ follows also from the
corresponding equations of motion which are trivially satisfyied.
In the papers \rF , \rZ
\ (see also \ref\rGT{M. P. Grabowski and C-H. Tze, {\it "On the octonionic
Nahm equations and self-dual membranes in 9-dimensions"}, Preprint
VPI--IHEP--92--2, Blacksburg, Julay, 1992.}) it was shown that the
action \eae \ has self-duality properties.  In the string case
these self-duality properties are easily demonstrated.  They
consist in the invariance of the action with respect to the
change:
$$
X^\mu _{,a}=
\epsilon ^{\mu \nu }\epsilon _{ab}\partial ^bX_{\nu }.
$$
We recall that the self-dual Yang-Mills theory is also invariant
with respect to the self-duality transformations.

The next step is to consider the action
with nondefinite sign
\eqn\eag{S=\kappa \int _{\cal M}d\tau d^p\sigma detX^\mu_{,a}.}
Incerting the identity
\eqn\eah{\eqalign {detX^{\mu }_{,a}
& ={1\over (p+1)!}\epsilon _{\mu ^0,\dots
,\mu ^p}\epsilon ^{a_0,\dots ,a_p}X^{\mu _0}_{a^0}\dots
X^{\mu _p}_{a^p} \cr
& =(-)^{p+1}\epsilon _{\mu ^0,\mu ^1,\dots ,\mu ^p}
\partial _{a^p}\biggl(X^{\mu _0}X^{\mu _1}_{,\tau }
X^{\mu _2}_{,\sigma ^1}\dots X^{\mu _p}_{,\sigma ^{p-1}}\biggr)\cr}}
in the action \eag \ and integrating over the
$p$-th direction we obtain (in the membrane case):
\eqn\eai{S=\kappa \int _{\cal \partial M}d\tau d\sigma \epsilon _{\mu \nu
\lambda }X^{\mu }X^{\nu }_\tau X^{\lambda }_{\sigma },}
where $\partial {\cal M }$ is the boundary of $\cal M $.
 Our result is that starting from the
topological $(p+1)$-brane we obtain C-S action for
$p$-dimensional extended object living in $p+2$-dimensional
space-time. For example the C-S particles are moving in
2-dimensional space-time, while the C-S strings are moving in
3-dimensional space time, etc..

The action \eai \ has simple geometrical meaning. It is just the area
momentum. In the particle case \eai \ gives area velocity.

In order to generalize the C-S objects to
any space-time dimension (as ordinary particles, strings, etc. )
we introduce the following notation:
$$X_{,A}=\partial _AX=\matrix{
 X      &  {\rm if}  & A=* \cr
 X_{,a} &  {\rm if}  & A=a, \cr }$$
where $a=0,1,2,\dots ,p$.
Then the Lagrangian in the action \eai \ takes a similar form as
in the topological action \eag , i.e. $L=detX^\mu_{,A}$ and
consequently, to generalize this action for target space with
arbitrary dimension, we introduce a generalized induced metric
tensor
$$
\widetilde g_{AB}=X^\mu _{,A}X^\nu _{,B}\eta _{\mu \nu }.
$$
By means of this metric tensor the form of the action for the C-S
p-brane moving in a target space with arbitrary dimension is the
same as the ordinary Nambu-Goto action, i.e.
\eqn\eal{S=
\kappa \int _{\cal {\partial M}}d\tau d^p\sigma
\sqrt{-\widetilde g},}

It is easy to check that the action \eai \ as well as \eal \
are invariant with respect to $p$-variable (if a start from
ordinary (p+1)-brane) diffeomorphism. As a concequences of these
invariance we have the following first class primary constraints:
\eqn\eca{\eqalign{
\phi _\perp &=
{\cal P}^2+\kappa ^2det\Bigl(X_{,u}X_{,v}\Bigr)\approx 0, \cr
\phi _j &={\cal P}X_{\sigma _j}\approx 0, \cr
\phi _* &={\cal P}X\approx 0,\cr }}
where $u,v=*,1,\dots ,p; \ j=1, \dots ,p$.  We note, that the
Lagrangian \eal \ and the constraints for the C-S $p$-brane \eca
\ can be obtained from the corresponding ordinary $(p+1)$-brane
Lagrangian \eaa \ and $(p+1)$-brane constraints \eac \ by
replacing $\partial _{\sigma _{p+1}}X$ by $X$. We recall, that the
ordinary bosonic string has only two (bilinear) constraints while
the C-S string has three primary costraints, one of which is of
degree four.

The appearance of the constraint $\phi _*$ shows us that we have
some residual symmetry from the $p+2$-variable diffeomorfisms
under which is invariant the action \eaa .

For any  C-S $p$-brane the canonical Hamiltonian vanishes
identically, i.e.
$$
{\cal H}_0={\cal P}\dot X -{\cal L}\equiv 0,
$$
which is a property of the ordinary p-brane theory too.

As was mentioned above, in the case of nonpolynomial Lagrangian
\eal \ which lives in arbitrary target space dimension we can
choose to have only first class constraints \rSZa , \rSZ \ or to have
also second class constraints. In the present article we restrict
our considerations only to the first possibility, i.e. to
have only the first class constraints. We recall that in the
ordinary string case we have two first class constraints, while
for the C-S string there are two possibilities: the
first is when we have four first class constraints \rSZ \ and the
second is when we have an infinite set of first class constraints
\rSZa .  In the latter case not all of the constraints are
independent if we deal with finite dimensional target space
hence we have not dynamical degree of freedom, i. e. we have
topological theory. In that case the dynamical degrees of freedom
are possible only if we have infinite dimensional target space.

In what follows we analize the constraint algebra for
nopolynomial C-S membarne. In this case we have four first class
primary constraints \eca . From the Poisson bracket
$\{\phi _\perp ,\phi _*\}_{PB}$ we obtain a secondary coinstraint
$$
\psi _{\perp }=det\Bigl(X_{,u}X_{,v}\Bigr)\approx 0.
$$
The appearance of this secondary constraint is a consequence of
the fact that in the constraints \eca \ the coordinates $X^\mu $
without derivative are present.

Further, consider the Poison bracket
\eqn\ecc{\{\phi _{\perp }(f),\psi _{\perp }(h)\}_{PB}\approx
2\int d^2\sigma fh\biggl(\bigl({\cal P}^\mu \partial _{\sigma
_1}^2X_\mu \bigr)
\Psi^{11}_{\perp }
+2\bigl({\cal P}^\mu \partial _{\sigma _1}\partial _{\sigma _2}X_\mu \bigr)
\Psi ^{12}_{\perp }
+\bigl({\cal P}^\mu \partial _{\sigma _2}^2
X_\mu \bigr)\Psi^{22}_{\perp }\biggr),}
which shows us that here appear new secondary constraints. From \ecc \ it
follows that we have two possibilities to introduce new secondary
constraints. The first is

\noindent a)
\eqn\ecd{\eqalign{\Psi ^{11}_{\perp } & ={\partial det(\bar g)\over \partial
X_{\sigma _1^2}}=X^2X^2_{\sigma _2}-\bigl(XX_{\sigma _2}\bigr)^2\approx0,
\cr
\Psi ^{12}_{\perp } & ={\partial det(\bar g)\over
\partial X^\mu _{,\sigma _1}X_{\mu ,\sigma _2}}=
\bigl(XX_{,\sigma _1}\bigr)\bigl(XX_{\sigma_2}\bigr)-
X^2\bigl(X_{\sigma _1}X_{\sigma _2}\bigr)\approx 0, \cr
\Psi^{22}_{\perp }& ={\partial det(\bar g)\over \partial X_{,\sigma _2}^2}=
X^2\bigl(X_{\sigma _2}\bigr)^2-\bigl(XX_{\sigma _2}\bigr)^2\approx0. \cr }}
The second possibility for introdusing new secondary constraits
consist in:

\noindent b)
\eqn\ece{\eqalign{\Gamma ^{1,0} & ={\cal P}\partial _{\sigma _1}^2X\approx 0,
\cr
\Gamma ^{1,1} & ={\cal P}\partial _{\sigma _1}\partial _{\sigma _2}X
\approx 0, \cr
\Gamma ^{0,2} & ={\cal P}\partial _{\sigma _2}^2X\approx 0. \cr }}
First we shall consider the case a). In this case the Poisson
bracket of the secondary constraints \ecd \ with the constraint
$\phi _\perp $ also allows two posibilities. The first of
them a$_1$) is when as independent new secondary constraints we chose:
$$
X^2\approx 0,
\qquad \bigl(\partial _{\sigma _j}X\partial _{\sigma _k}X\bigr)\approx 0,
\qquad (j,k=1,2).
$$

In the second case a$_2$) the new secondary constaraints coinside
with \ece , i.e. with the case b). Consequently, for the C-S
membrane from both possible choises of the secondary constraints
by multiple application of the Poisson brackets we find an infinite
serie of secondary constraints. As one appropriate choice of the
secondary constraints is the following:
\eqn\ech{\eqalign{\Psi ^{m,n} & =\Bigl({\cal P}\partial _{\sigma _1}^m
\partial _{\sigma _2}^n{\cal P}\Bigr)\approx 0, \cr
\Phi ^{m,n} & =\Bigl(X\partial _{\sigma _1}^m\partial _{\sigma
_2}^nX\Bigr)\approx 0, \cr
\Gamma ^{m,n} & =\Bigl({\cal P}\partial _{\sigma 1}^m\partial _{\sigma
_2}^nX\Big)\approx 0, \qquad  (m,n=0,1,\dots ). \cr }}

We note that when the target space dimesion $D$ is finite
we have only a finite number of independent constraints \ech ,
hence, in that case we have a theory without dynamical
degree of freedom, i.e. we have a topological theory.
In the case when we are dealing with infinite dimensional target
space it is easy to check that idependent are only the
constraints $\Psi $ and $\Phi $ for which $m+n=2k$
($k=0,1,\dots $).

With respect to the Poisson brackets the constraints \ech \ form
the following infinite algebra:
\eqn\eci{\eqalign{
\{\Phi ^{k,l}\bigl[f\bigr],\Psi ^{m,n}\bigl[h\bigr]\}_{PB} & =
\sum _{r=0}^m\sum _{s=0}^n\pmatrix{m \cr r \cr}
\pmatrix{n \cr s \cr }\Biggl((-)^{m+n}\Phi ^{k+m-u,l+n-v}\bigl[
\partial _{\sigma _1}^u\partial _{\sigma _2}^v(fh)\bigr] \cr
& + \sum _{u=0}^{m-r}\sum _{v=0}^{n-s}(-){r+s}\pmatrix{m-r \cr u \cr }
\pmatrix{ n-s \cr v \cr } \cr
& \times \Phi ^{k+m-r-u,l+n-s-v}\bigl[
\partial _{\sigma _1}^u\partial _{\sigma _2}^v\bigl(
f\partial _{\sigma _1}^r\partial _{\sigma _2}^sh\bigr)\bigr]
 \cr
& (-)^{k+l+m+n}\sum _{u=0}^k\sum_{v=0}^l\pmatrix{k \cr u \cr }
\pmatrix{ l \cr v \cr } \cr
& \times \Phi ^{k+m-r-u,l+n-s-v}\bigl[
\partial _{\sigma _1}^r\partial _{\sigma _2}^s\bigl(h
\partial _{\sigma _1}^u\partial _{\sigma _2}^v\bigr)\bigr] \cr
& (-)^{k+l+r+s}\sum _{p=0}^k\sum _{q=0}^l\sum _{u=0}^{m-r}
\sum _{v=0}^{n-s}\pmatrix{k \cr p \cr }\pmatrix{l \cr q \cr }
\pmatrix{m-r \cr u \cr }\pmatrix{n-s \cr v \cr } \cr
& \times \Phi ^{k+m-r-p-u,l+n-s-q-v}\bigl[\partial _{\sigma _1}^u
\partial _{\sigma _2}^v\bigl(\partial _{\sigma _1}^p
\partial _{\sigma _2}^qf\partial _{\sigma _1}^r
\partial _{\sigma _2}^sh\bigr)\bigr]\Biggr), \cr }}
\eqn\ecj{\eqalign{
\{\Gamma ^{k,l}\bigl[f\bigr],\Psi ^{m,n}\bigl[h\bigr]\}_{PB} & =
\sum _{p=0}^k\sum _{q=0}^l\sum _{u=0}^{k-p}\sum _{v+0}^{l-q}
\pmatrix{k \cr p \cr }\pmatrix{ l \cr q \cr }
\pmatrix{k-p \cr u \cr }\pmatrix{l-q \cr v \cr } \cr
 & \times (-)^{p+q}\Biggl(\Psi ^{k+m-p-u,l+n-q-v}\bigl[
\partial _{\sigma _1}^u\partial _{\sigma _2}^v\bigl(
h\partial _{\sigma _1}^p\partial _{\sigma _2}^qf\bigr)\bigr]
 \cr
& (-)^{m+n}\sum _{r=0}^m\sum_{s=0}^n\pmatrix{m \cr r \cr }
\pmatrix{ n \cr s \cr } \cr
 & \times \Psi ^{k+m-p-r-u,l+n-q-s-v}\bigl[
\partial _{\sigma _1}^u
\partial _{\sigma _2}^v\bigl(\partial _{\sigma _1}^p
\partial _{\sigma _2}^qf\partial _{\sigma _1}^r
\partial _{\sigma _2}^sh\bigr)\bigr]\Biggr), \cr }}
\eqn\eck{\eqalign{
\{\Phi ^{k,l}\bigl[f\bigr],\Gamma ^{m,n}\bigl[h\bigr]\}_{PB} & =
\sum _{p=0}^k\sum _{q=0}^l
\pmatrix{k \cr p \cr }\pmatrix{ l \cr q \cr }
\Biggl((-)^{k+l}\Phi ^{k+m-p,l+n-q}\bigl[
\partial _{\sigma _1}^p\partial _{\sigma _2}^q\bigl(fh\bigr)\bigr]
 \cr
& +\sum _{u=0}^{k-p}\sum _{v+0}^{l-q}(-)^{p+q}
 \pmatrix{k-p \cr u \cr }\pmatrix{l-q \cr v \cr } \cr
& \times \Phi ^{k+m-p-u,l+n-q-v}\bigl[\partial _{\sigma _1}^u
\partial _{\sigma _2}^v\bigl(h\partial _{\sigma _1}^p
\partial _{\sigma _2}^qf\bigr)\bigr]\Biggr), \cr }}
\eqn\ecl{\eqalign{
\{\Gamma ^{k,l}\bigl[f\bigr],\Gamma ^{m,n}\bigl[h\bigr]\}_{PB} & =
\sum _{p=0}^k\sum _{q=0}^l\sum _{u=0}^{k-p}\sum _{v=0}^{l-q}
(-)^{u+v}\pmatrix{k \cr p \cr}\pmatrix{l \cr q \cr }
\pmatrix{k-p \cr u \cr }\pmatrix{ l-q \cr v \cr } \cr
& \times \Gamma ^{k+m-p-u,l+n-q-v}\bigl[
\partial _{\sigma _1}^{k+u-p}\partial _{\sigma _2}^{l+v-q}
\bigl(
h\partial _{\sigma _1}^p\partial _{\sigma _2}^qf\bigr)\bigr] \cr
& -\sum _{p=0}^m\sum _{q=0}^n\sum _{u=0}^{m-p}\sum _{v=0}^{n-q}
(-)^{u+v}\pmatrix{m \cr p \cr}\pmatrix{n \cr q \cr }
\pmatrix{m-p \cr u \cr }\pmatrix{ n-q \cr v \cr } \cr
& \times \Gamma ^{k+m-p-u,l+n-q-v}\bigl[
\partial _{\sigma _1}^{m+u-p}\partial _{\sigma _2}^{n+v-q}
\bigl(
f\partial _{\sigma _1}^p\partial _{\sigma _2}^qh\bigr)\bigr]. \cr }}

Now we are looking for subalgebras of this algebra. It is easy to see
that the constraints $\Phi ^{0,0}, \ \Psi ^{0,0} \ {\rm and } \ \Gamma
^{0,0}$ satisfy the $SL(2,R)$ Kac-Moody algebra. Indeed,
introducing notations
\eqn\ecm{H={\cal P}X\approx 0, \qquad
E_+={\cal P}^2\approx 0, \qquad
E_-=X^2\approx 0,}
we have just the generators of the deformed $SL(2,R)$ Kac-Moody algebra:
\eqn\ecma{\eqalign{\{H[f],E_\pm [h]\}_{PB} & =\pm 2E_\pm [fh], \cr
\{E_+[f],E_-[h]\}_{PB} & =\lambda H[fh], \cr }}
where the parameter of deformation $\lambda =-4$.

The constraints $\Gamma _j={\cal P}\partial _{\sigma _j}X, (j=1,2)$ satisfy
two mutually noncommuting\foot{This is a difference with the
Virasoro algebras in the ordinary string theory.} Virasoro algebras
\eqn\ecmb{\{\Gamma ^j[f],\Gamma ^k[h]\}_{PB}=
\Gamma _k\bigl[f\partial _{\sigma _j}h\bigr]
-\Gamma _k\bigl[h\partial _{\sigma _j}f\bigl].}

It is easy to see that any of these Virasoro algebras can be
extended to $W_\infty $ algebras. More strictly, we have two Virasoro
and two $W_{1+\infty }$ algebras acting on the $\sigma _j$ direction
complemented by rotations in the $\sigma _1, \sigma _2$ plane.

The $SL(2,R)$ Kac-Moody generators and the Virasoro generators
also form an subalgebra including \ecma , \ecmb \ and which is
closed on
\eqn\ecmc{\eqalign{\{H[f],\Gamma ^j[h]\}_{PB} & =
H[f\partial _{\sigma _j}]-H[\partial _{\sigma _j}(fh)], \cr
\{E_+[f],\Gamma ^j[h]\}_{PB} & =
E_+[f\partial _{\sigma _j}h-h\partial _{\sigma _j}f], \cr
\{E_-[f],\Gamma ^j[h]\}_{PB} & =-E_-[\partial _{\sigma _j}(fh)]. \cr }}

Concequently, we have as subalgebras also two copies $W_{1+\infty
}$-algebras in linear realization\foot{As the Virasoro
subalgebras, these $W_{1+\infty }$-algebras differ from the
ordinary $W_{1+\infty }$-algebras because they are not mutually
commuting.}. One of them acts on $\sigma _1$ coordinate and the
second acts on the $\sigma _2$ coordinate. As generators of
these transformations appear the constraints $\Gamma ^{k,0} \ {\rm
and} \ \Gamma ^{0,k}$.  In the general case $\Gamma ^{k,l}$ can be
considered as generators of generalized diffeomorphismes in two
dimensional space. Indeed, the Poisson bracket of $\Gamma $ with
the coordinate $X^\mu \ {\rm and} \ {\cal P}^\mu $ give the
transformation laws for the phase space coordinates:
\eqn\ecn{\eqalign{\delta _\Gamma ^{k,l}X^\mu =
\{\Gamma ^{k,l}\bigl[f\bigr],X^\mu \}_{PB} & =
-f\partial _{\sigma _1}^k\partial _{\sigma _2}^lX^\mu ,\cr
\delta _\Gamma ^{k,l}{\cal P}^\mu =
\{\Gamma ^{k,l}\bigl[f\bigr],X^\mu \}_{PB} & =
(-)^{k+l}\sum _{p=0}^k\sum _{q=0}^l\pmatrix {k \cr p \cr }
\pmatrix {l \cr q \cr }\partial _{\sigma _1}^p\partial _{\sigma
_2}^qf\partial _{\sigma _1}^{k-p}\partial _{\sigma_2}^{l-q}
{\cal P}^\mu .\cr }}
In the same way we obtain also:
\eqn\eco{\eqalign{\delta _{\Phi }^{k,l}X^{\mu } & =
\{\Phi ^{k,l}\bigl[f\bigr],X^{\mu }\}_{PB}=0, \cr
\delta _{\Phi }^{k,l}{\cal P}^{\mu } & =
\{\Phi ^{k,l}\bigl[f\bigr],{\cal P}^{\mu }\}_{PB} \cr
& =-f\partial _{\sigma _1}^{k}
\partial _{\sigma _2}^{l}X^{\mu }
-(-)^{k+l}\sum _{p=0}^{k}\sum _{q=0}^{l}
\pmatrix{k \cr p \cr }\pmatrix{l \cr q \cr }
\partial _{\sigma _1}^{p}\partial _{\sigma _2}^{q}f\partial _{\sigma _1}^{k-p}
\partial _{\sigma _2}^{l-q}X^{\mu }, \cr
\delta _{\Psi }^{k,l}X^{\mu } & =
\{\Psi ^{k,l}\bigl[f\bigr],X^{\mu }\}_{PB}
=f\partial _{\sigma _1}^{k}
\partial _{\sigma _2}^l{\cal P}^{\mu }(-)^{k+l}\sum _{p=0}^k\sum _{q=0}^l
\pmatrix{k \cr p \cr }\pmatrix{l \cr q \cr }
\partial _{\sigma _1}^{p}\partial _{\sigma _2}^{q}f\partial _{\sigma _1}^{k-p}
\partial _{\sigma _2}^{l-q}{\cal P}^{\mu }, \cr
\delta _{\Psi }^{k,l}{\cal P}^{\mu } & =
\{\Phi ^{k,l}\bigl[f\bigr],{\cal P}^{\mu }\}_{PB}=0, \cr }}
Consequently, $\delta _\Gamma ^{1,0} \ {\rm and} \ \delta _\Gamma
^{0,1}$ are the ordinary diffeomorphisms in two-dimensional
space.  We note that the appearing assymmetry of transformation
laws of the coordinate $X$ and momentum ${\cal P}$ is a
consequence of the assymetric choise of the constraints \ech .
Taking into account the identity
$$
\partial ^m{\cal P}\partial ^nX=\sum_{q=0}^m(-)^q\pmatrix{m \cr q
\cr }\partial ^q\bigl({\cal P}\partial ^{m+n-q}X\bigr)
$$
it follows that the more symmetric basis can be obtained for the
constraints \ech \ by a simple redefinition
$$
\Lambda ^{m,n}\rightarrow \widetilde \Lambda ^{m,n}=
\sum _{p=0}^m\sum _{q=0}^{n}b^{mn}_{pq}\partial _{\sigma _1}^p
\partial _{\sigma _2}^{q}\Lambda ^{m-p,n-q},
$$
where $b$ are some constants. By a suitable choise
of the constants $b$ the classical algebra \eci --\ecl \ can be
deformed to the algebra which admits diagonal central extension
\rP \  at least for $W_{1+\infty }$ subalgebra.

In order to determine the conformal spin content of the
constraints \ech
\ we consider their conformal transformation laws:
\eqn\ecq{\eqalign{\delta _\Gamma ^{1,0}\bigl[f\bigr]\Gamma ^{m,n} &=
(m+1)\partial _{\sigma _1}f\Gamma ^{m,n}+
f\partial _{\sigma _1}\Gamma ^{m,n}
+\sum _{p=2}^m\sum _{q=0}^n\pmatrix{m \cr p \cr }\pmatrix{n \cr
q \cr }\partial _{\sigma _1}^p\partial _{\sigma _2}^q\Gamma
^{m-p+1,n-q}, \cr
\delta _\Gamma ^{0,1}\bigl[f\bigr]\Gamma ^{m,n} &=
(n+1)\partial _{\sigma _2}f\Gamma ^{m,n}+
f\partial _{\sigma _2}\Gamma ^{m,n}
+\sum _{p=0}^m\sum _{q=2}^n\pmatrix{m \cr p \cr }\pmatrix{n \cr
q \cr }\partial _{\sigma _1}^p\partial _{\sigma _2}^q\Gamma
^{m-p,n-q+1}, \cr
\delta _\Gamma ^{1,0}\bigl[f\bigr]\Psi ^{m,n} &=
(m+2)\partial _{\sigma _1}f\Psi ^{m,n}+
f\partial _{\sigma _1}\Psi ^{m,n}
+\sum _{p=1}^m\sum _{q=0}^n\pmatrix{m \cr p \cr }\pmatrix{n \cr
q \cr }\partial _{\sigma _1}^{p+1}\partial _{\sigma _2}^q\Psi
^{m-p,n-q}, \cr
& +\sum _{p=2}^m\sum _{q=0}^n\pmatrix{m \cr p \cr }\pmatrix{n \cr
q \cr }\partial _{\sigma _1}^{p}\partial _{\sigma _2}^q\Psi
^{m-p+1,n-q}, \cr
\delta _\Gamma ^{0,1}\bigl[f\bigr]\Psi ^{m,n} &=
(m+2)\partial _{\sigma _2}f\Psi ^{m,n}+
f\partial _{\sigma _2}\Psi ^{m,n}
+\sum _{p=0}^m\sum _{q=1}^n\pmatrix{m \cr p \cr }\pmatrix{n \cr
q \cr }\partial _{\sigma _1}^{p}\partial _{\sigma _2}^{q+1}\Psi
^{m-p,n-q}, \cr
& +\sum _{p=0}^m\sum _{q=2}^n\pmatrix{m \cr p \cr }\pmatrix{n \cr
q \cr }\partial _{\sigma _1}^{p}\partial _{\sigma _2}^q\Psi
^{m-p,n-q+1}, \cr
\delta _\Gamma ^{1,0}\bigl[f\bigr]\Phi ^{m,n} &=
m\partial _{\sigma _1}f\Phi ^{m,n}+
f\partial _{\sigma _1}\Phi ^{m,n}
+\sum _{p=2}^m\sum _{q=0}^n\pmatrix{m \cr p \cr }\pmatrix{n \cr
q \cr }\partial _{\sigma _1}^{p}\partial _{\sigma _2}^q\Phi
^{m-p+1,n-q}, \cr
\delta _\Gamma ^{0,1}\bigl[f\bigr]\Phi ^{m,n} &=
m\partial _{\sigma _2}f\Phi ^{m,n}+
f\partial _{\sigma _2}\Phi ^{m,n}
+\sum _{p=0}^m\sum _{q=2}^n\pmatrix{m \cr p \cr }\pmatrix{n \cr
q \cr }\partial _{\sigma _1}^{p}\partial _{\sigma _2}^q\Phi
^{m-p,n-q+1}. \cr }}
{}From \ecq \ it follows that the constraint $\Phi ^{m,n}$ has
conformal spin $s_1=m$ in the $\sigma _1$ direction and $s_2=n$
in the $\sigma _2$ direction. The constraints $\Psi \ {\rm and}
\ \Gamma $ have spins $s_1=m+1, \ s_2=n+1 \ {\rm and} \ s_1=m+2,
 \ s_2=n+2$ respectively. From \ecq \ it follows also that only the
constraints $\Gamma ^{0,0}, \ \Gamma ^{1,0}, \ \Gamma ^{0,1}, \ \Phi
^{0,0}, \ \Phi ^{1,0}, \ \Phi ^{0,1} \ {\rm and} \ \Psi ^{0,0}$ are
transformed  according to the primary field conformal law.

We note that considerations presented here can be generalized
for arbitrary $p$-branes without difficulties. In such a way we find a
linear realization of higher spin extension of the
$p$-dimensional $SL(2,C)$ Kac-Moody algebra containing as
subalgebras $p$ copies mutually noncommuting $W_{1+\infty }$
algebras. We note, that in the considered above C-S membrane
case $(p=2)$ we have linear realization of higher spin extension
of two-dimensional $SL(2.R)$ Kac-Moody algebra containing as
subalgebras two (mutually  noncommuting) $W_{1+\infty }$ algebras
which differ from the ordinary two-dimensional conformal
transformations in the following: 1) The transformations act only
on the spatial coordinates while the evolution parameter is not
attacked. The total algebra is not represented as a direct product
of two mutualy commuting algebras each of which acts in
one-dimensional space. The latter shows us that we have a
nontrivial extension of the higher spin symmetry algebras  acting
in space-time of arbitrary dimension. Moreover, our
considerations differ by those in \rMR \ and \rMMR \ because they
are connected with a field theorethical model and we have a
linear realization of higher spin extended $p$-dimensional
$SL(2,R)$ Kac-Moody algebra acting on the phase-space for the C-S
$p$-brane.

\listrefs
\bye